\documentstyle[12pt]{article}
\makeatletter

\newdimen\LENB \newdimen\LENW \newdimen\THI
\newdimen\LENWH \newdimen\LENTOT \newcount\N
\def\vbrknlnele#1#2#3{
  \LENB=#1pt \LENW=#2pt \THI=#3pt
  \LENWH=\LENW \divide\LENWH by 2
  \LENTOT=\LENB \advance\LENTOT by \LENW
  \vbox to \LENTOT{
    \vbox to \LENWH{}
    \nointerlineskip
    \vbox to \LENB{\hbox to \THI{\vrule width \THI height \LENB}}
    \nointerlineskip
    \vbox to \LENWH{}
  }}

\def\vbrknln#1{
  \N=#1
  \vcenter{
    \vbox{
      \loop\ifnum\N>0
        \vbox to 4pt{\vbrknlnele{2}{2}{0.1}}
        \nointerlineskip
        \advance\N by -1
      \repeat
  }}}

\def\vbl#1{\hskip-5pt \vbrknln{#1} \hskip-5pt}

\def\hbrknlnele#1#2#3{
  \LENB=#1pt \LENW=#2pt \THI=#3pt
  \LENTOT=\LENB \advance\LENTOT by \LENW
  \vcenter{
    \vbox to \THI{
      \hbox to \LENTOT{
        \hfil
        \vrule width \LENB height \THI
        \hfil}
  }}}

\def\hblele{\hbrknlnele{2}{2.2}{0.1}}

\def\hblfil{\cleaders\hbox{$ \m@th \mkern1mu \hblele \mkern1mu
$}\hfill}

\def\eqnarray{%
\stepcounter{equation}%
\let\@currentlabel=\theequation
\global\@eqnswtrue
\global\@eqcnt\z@
\tabskip\@centering
\let\\=\@eqncr
$$\halign to \displaywidth\bgroup\@eqnsel\hskip\@centering
$\displaystyle\tabskip\z@{##}$&\global\@eqcnt\@ne
\hfil$\displaystyle{{}##{}}$\hfil
&\global\@eqcnt\tw@$\displaystyle\tabskip\z@{##}$\hfil
\tabskip\@centering&\llap{##}\tabskip\z@\cr}

\def\@cite#1#2{\unskip\nobreak\relax
     \def\@tempa{$\m@th{\hbox{[#1]}}$}%
    \futurelet\@tempc\@citexx}

\def\@citexx{\ifx.\let\@tempd=\@citepunct \@tempc\else
    \ifx,\let\@tempd=\@citepunct \@tempc\else
    \let\@tempd=\@tempa\fi\fi\@tempd}
\def\@citepunct{\@tempc\edef\@sf{\spacefactor=\the\spacefactor\relax}\@tempa
    \@sf\@gobble}

\def\citenum#1{{\def\@cite##1##2{##1}\cite{#1}}}
\def\citea#1{\@cite{#1}{}}

\newcount\@tempcntc
\def\@citex[#1]#2{\if@filesw\immediate\write\@auxout{\string\citation{#2}}\fi
  \@tempcnta\z@\@tempcntb\m@ne\def\@citea{}\@cite{\@for\@citeb:=#2\do
    {\@ifundefined
       {b@\@citeb}{\@citeo\@tempcntb\m@ne\@citea\def\@citea{,}{\bf ?}\@warning
       {Citation `\@citeb' on page \thepage \space undefined}}%
    {\setbox\z@\hbox{\global\@tempcntc0\csname b@\@citeb\endcsname\relax}%
     \ifnum\@tempcntc=\z@ \@citeo\@tempcntb\m@ne
       \@citea\def\@citea{,}\hbox{\csname b@\@citeb\endcsname}%
     \else
      \advance\@tempcntb\@ne
      \ifnum\@tempcntb=\@tempcntc
      \else\advance\@tempcntb\m@ne\@citeo
      \@tempcnta\@tempcntc\@tempcntb\@tempcntc\fi\fi}}\@citeo}{#1}}
\def\@citeo{\ifnum\@tempcnta>\@tempcntb\else\@citea\def\@citea{,}%
  \ifnum\@tempcnta=\@tempcntb\the\@tempcnta\else
   {\advance\@tempcnta\@ne\ifnum\@tempcnta=\@tempcntb \else \def\@citea{--}\fi
    \advance\@tempcnta\m@ne\the\@tempcnta\@citea\the\@tempcntb}\fi\fi}

\makeatother
\def\romanno#1{\uppercase\expandafter{\romannumeral#1}}
\def\ld{\lambda}
\def\dis{\displaystyle}

\begin{document}

\begin{titlepage}

\begin{center}

\begin{Large}
{\bf Casorati Determinant Solutions for the Discrete}\\[2mm]
{\bf Painlev\'e III Equation}\\[3mm]
\end{Large}

\vspace{30pt}

\begin{normalsize}
{Kenji Kajiwara}\\
{\it Department of Electrical Engineering,
Doshisha University,}\\
{\it Tanabe, Kyoto 610-03, Japan}\\
{Yasuhiro Ohta}\\
{\it Department of Applied Mathematics, Faculty of Engineering,}\\
{\it Hiroshima University, }\\
{\it 1-4-1 Kagamiyama, Higashi-Hiroshima, Hiroshima 724, Japan}\\
{and}\\
{Junkichi Satsuma}\\
{\it Department of Mathematical Sciences, University of Tokyo,}\\
{\it 3-8-1 Komaba, Meguro-ku, Tokyo 153, Japan}
\end{normalsize}
\end{center}
\vspace{30pt}
\begin{abstract}
The discrete Painlev\'e III equation is investigated based on the bilinear
formalism. It is shown that it admits
the solutions expressed by the Casorati determinant
whose entries are given by the discrete
Bessel function. Moreover,
based on the observation that these discrete Bessel functions are
transformed to the $q$-Bessel functions by a simple variable transformation,
we present a $q$-difference analogue of the Painlev\'e III equation.
\end{abstract}

\end{titlepage}

\addtolength{\baselineskip}{.3\baselineskip}

\section{Introduction}
Recently, several interesting ordinary difference equations
have been proposed as discrete analogues of the Painlev\'e equations.
Among them, the discrete Painlev\'e I(dP$_{\rm I}$) and II(dP$_{\rm II}$)
are naturally derived in calculating the partition functions
of some models in the theory of two-dimensional quantum
gravity\cite{Brezin,Periwal,Its}.
In other words, they have close relationship with the theory of
orthogonal polynomials. This example implies the significance
of discrete Painlev\'e equations. Moreover,
they are expected to possess rich mathematical structures.
Besides this derivation, the discrete Painlev\'e equations
can be obtained in several context\cite{dMKdV,dP,dPreview1,dPreview2,SC}.

It is known that the discrete Painlev\'e equations admit
several properties analogous to continuous ones,
such as the existence of Lax pair, B\"acklund transformation,
coalescence cascade and singularity confinement(SC) property
\cite{dP,dPreview1,dPreview2,Iso,Miura,dP4,dP3}.
However, are these properties sufficient to regard
the equations really as the discrete Painlev\'e equations?

In the continuous case, the Painlev\'e equations are
``integrated" in the sense that they are reduced to
linear integral equations by means of inverse scattering
transform from their Lax pairs\cite{ARS}.
In the discrete case, however, the dP$_{\rm I}$ equation is the only
known example which is ``integrated" based on its Lax pair.

In the continuous case, series of exact solutions which
are expressed by determinants are obtained by using
the B\"acklund transformation\cite{Okamoto}. However,
no attempt has been done to construct such solutions for the discrete
case.

The coalescence cascade is a property of the Painlev\'e equations
which is related to the
confluence of singularities
of linear ordinary differential equations in the theory
of monodromy preserving deformation. Hence, this property
is considered to have no direct relationship to the structure of
their solution spaces.

The SC is a concept which was extracted as a common property
{}from the class of integrable ordinary difference equations proposed by
Quispel et al.\cite{Quispel}\  In this context, ``integrable" means the
possession
of conserved quantities. Grammaticos et al. have employed
this property as an integrability criterion of difference
equations\cite{SC}. By applying this test to non-autonomous equations, they
have proposed several discrete Painlev\'e equations\cite{dP}.
Since the SC might be regarded as the discrete analogue of the Painlev\'e
property, it provides us with a useful and powerful tool to detect
``good" equations. However, we do not know whether the equations
which pass the SC are really ``integrable" in the above sense.

The Painlev\'e equations are characterized by the property
that they do not possess movable singularities except for poles.
It is also a significant property that
they admit series of exact solutions expressed by
determinants for special values of parameters, although
their solutions are transcendental in general.
We believe that the existence of such solutions is a direct
reflection of the ``integrability" of equations.
{}From this viewpoint, we consider that the preservation
of the structure of solution space is the most important
for the discretization of Painlev\'e equations.

In the previous paper, we have shown that the dP$_{\rm II}$ admits
a class of solutions expressed by the Casorati determinants
of discrete Airy function\cite{dP2}.
They reduce to the Wronskian solutions of P$_{\rm II}$
in the continuous limit.
In this paper, we investigate the solutions of special function
type for the discrete Painlev\'e III (dP$_{\rm III}$) equation.
We will show that they are expressed in terms of the Casorati
determinants of the discrete Bessel functions.
The dP$_{\rm III}$ occupies a particular place among the discrete
Painlev\'e equations, since this is the simplest equation
which is derived only through the SC. Hence, the existence of such
solutions may be a good check of the validity of SC.

Moreover, we will show that the discrete Bessel functions in
our result are transformed to the $q$-Bessel functions
by a simple independent variable transformation.
Based on this observation, we present a $q$-difference
analogue of the P$_{\rm III}$.

In section 2, we present the explicit form of exact solutions for
the dP$_{\rm III}$. The derivation is discussed in detail by
means of the bilinear formalism in section 3.
In section 4, we present a $q$-difference analogue of the P$_{\rm III}$.
Finally in section 5, we give concluding remarks.
\section{Solutions of dP$_{\rm\romanno3}$}
In this section, we investigate the solutions of dP$_{\rm\romanno3}$\cite{dP},
\begin{equation}
w_{n+1}w_{n-1}=\frac{\alpha w_n^2
+\beta \lambda^{n} w_n+\gamma \lambda^{2n}
}{w_n^2+\delta w_n+\alpha}\ ,\label{dP3-1}
\end{equation}
where $\alpha$, $\beta$, $\gamma$, $\delta$, and $\lambda$ are parameters.
Before going to the discrete case, let us remind of the special
function type solutions for the P$_{\rm III}$ equation,
\begin{equation}
\frac{d^2u}{dx^2}=\frac{1}{u}
\left( \frac{du}{dx}\right)^2-\frac{1}{x}\frac{du}{dx}
+\frac{1}{x}(\alpha u^2+\beta)
+ \gamma u^3+\frac{\delta}{u}\ ,\label{P3}
\end{equation}
where $\alpha$, $\beta$, $\gamma$ and $\delta$ are
parameters. It is known that eq.(\ref{P3}) admits the
solutions expressed by the $\tau$ function\cite{Okamoto},
\begin{equation}
\tau_N^{\nu}=
\left\vert\matrix{
J_{\nu} & D~J_{\nu} & \cdots & D^{N-1}J_{\nu}\cr
D~J_{\nu} & D^2~J_{\nu} & \cdots & D^{N}~J_{\nu}\cr
\vdots    &\vdots      &\ddots   &\vdots\cr
D^{N-1}J_{\nu} & D^N~J_{\nu} & \cdots & D^{2N-2}~J_{\nu}\cr}\right\vert
\ ,\label{P3tau}
\end{equation}
where $D=x{\displaystyle d\over\displaystyle dx}$ and $J_{\nu}$
is the Bessel function of degree $\nu$. The dependent variable $u$
in eq.(\ref{P3}) is expressed as
\begin{equation}
u = \biggl( {\rm log}~{\displaystyle \tau^{\nu+1}_N\over\displaystyle
\tau^{\nu}_{N+1}}\biggr)_x + {\displaystyle \nu+N\over\displaystyle
x} \ .\label{P3var}
\end{equation}
Equation (\ref{P3var}) gives the particular solutions of eq.(\ref{P3})
with the parameters,
\begin{equation}
\alpha=2(N-\nu)\ ,\quad\beta=2(\nu+N+1)\ ,\quad\gamma=1\ ,
\quad \delta=-1\ .\label{P3par}
\end{equation}

To obtain the particular solutions of dP$_{\rm III}$ eq.(\ref{dP3-1}), we
start from the discrete Riccati equation \cite{dRiccati}.
In this case, it is easily shown that if $w_n$ satisfies
\begin{equation}
w_{n+1}=-\frac{aw_n+\lambda^n}{w_n+d}\ ,
\label{Riccatio}
\end{equation}
where $a$ and $d$ are constants,
then $w_n$ gives a solution of eq.(\ref{dP3-1}).
Assuming the fractional form,
\begin{equation}
w_n=\frac{F_n}{G_n}\ ,
\end{equation}
and substituting this into eq.(\ref{Riccatio}),
we get
\begin{eqnarray}
&w_n=\frac{G_{n+1}}{G_n} - d\ ,&\label{fraco}\\
&G_{n+2}+(a-d)G_{n+1}+(\lambda^n-ad)G_n=0\ .&\label{dBessel1}
\end{eqnarray}
On the analogy of the continuous case, it is expected that
eq.(\ref{dBessel1}) is a discrete version of the Bessel equation. However,
it is not trivial to identify eq.(\ref{dBessel1}) with the Bessel equation.
Instead of eqs.(\ref{fraco}) and (\ref{dBessel1}), we consider
\begin{equation}
w_n=\frac{J_{\nu}(n+1)}{J_{\nu}(n)}-q^{\nu},
\label{fracJo}
\end{equation}
\begin{equation}
J_{\nu}(n+2)-(q^{\nu}+q^{-\nu})J_{\nu}(n+1)+\{ 1+(1-q)^2q^{2n}\}
J_{\nu}(n)=0\ .\label{dBessel2}
\end{equation}
We can show that if $J_{\nu}(n)$ satisfies
eq.(\ref{dBessel2}), then $w_n$ in eq.(\ref{fracJo})
gives the solution of dP$_{\rm\romanno3}$ eq.(\ref{dP3-1}) with the parameters,
\begin{equation}
\alpha=-1,\quad \beta=(q^{\nu}-q^{-\nu-2})(1-q)^2,\quad \gamma=
\frac{(1-q)^4}{q^2},\quad \delta=q^{\nu}-q^{-\nu},\quad \lambda=q^2\ .
\end{equation}
Continuous limit of eq.(\ref{dBessel2}) is given by putting
$q=1+\epsilon$, $n=\frac{\displaystyle z}{\displaystyle\epsilon}$
and taking the limit
of $\epsilon\rightarrow 0$, which yields
\begin{equation}
\frac{d^2J_{\nu}}{dz^2}+({\rm e}^{2z}-\nu^2)J_{\nu}=0\ .
\label{Bessel1}
\end{equation}
Equation (\ref{Bessel1}) reduces to the Bessel equation in ordinary form
if we put $x={\rm e}^z$.
In this sense, eq.(\ref{dBessel2})
is regarded as the discrete Bessel equation. We note that
the same results for this simplest solution is also derived
in ref.\cite{dP3}.

The critical point is that the above result can be extended to
the $N\times N$ determinant solution.
We here give only the results, leaving
the derivation in the next section.
We consider the $\tau$ function,
\begin{equation}
\tau_N^{\nu}(n)=\left\vert\matrix{
J_{\nu}(n)     &J_{\nu}(n+1)  &\cdots &J_{\nu}(n+N-1) \cr
J_{\nu}(n+2)   &J_{\nu}(n+3)  &\cdots &J_{\nu}(n+N+1)\cr
\vdots         &\vdots &\cdots&\vdots\cr
J_{\nu}(n+2N-2)&J_{\nu}(n+2N-1)&\cdots &J_{\nu}(n+3N-3)\cr}\right\vert\ ,
\label{dP3tau}
\end{equation}
where $J_{\nu}(n)$ satisfies the following contiguity relations,
\begin{eqnarray}
 J_{\nu}(n)-q^{-\nu}J_{\nu}(n+1)&=&-(1-q)q^nJ_{\nu+1}(n)\ ,
\label{contig1}\\
 J_{\nu}(n)-q^{\nu}J_{\nu}(n+1)&=&(1-q)q^nJ_{\nu-1}(n)\ .\label{contig2}
\end{eqnarray}
We notice that the discrete Bessel equation (\ref{dBessel2}) is derived
>from above contiguity relations.
Then we can show that
\begin{equation}
w_n=\frac{\tau_{N+1}^{\nu}(n+1)\tau_N^{\nu+1}(n)}
{\tau_{N+1}^{\nu}(n)\tau_N^{\nu+1}(n+1)}-q^{\nu+N}\ ,
\label{dP3var}
\end{equation}
gives the solutions of the
dP$_{\rm\romanno3}$,
\begin{equation}
w_{n+1}w_{n-1}=\frac{\alpha w_n^2
+\beta q^{2n} w_n+\gamma q^{4n}}{w_n^2+\delta w_n+\alpha},
\label{dP3-2}
\end{equation}
with the parameters,
\begin{equation}
\begin{array}{ll}
\alpha=-q^{4N}\ ,&\beta=(q^{\nu+N}-q^{-\nu-N-2})~q^{8N}(1-q)^2\ ,
\\
\gamma=q^{2(6N-1)}~(1-q)^4\ ,&\delta=(q^{\nu-N}-q^{-\nu+N})~q^{2N}\ .
\end{array}
\label{dP3par}
\end{equation}
It is remarkable that the $\tau$ function has a similar asymmetric
structure to that
of dP$_{\rm\romanno2}$\cite{dP2}. Namely,
the index of elements of the determinant, $n$, increases by two in one
direction while by one in another direction.

\section{Derivation of Results}

In this section, we show that the $\tau$ function (\ref{dP3tau}) actually
gives the solution of eq.(\ref{dP3-2}) with the parameters
({\ref{dP3par}),
through the dependent variable transformation (\ref{dP3var}).
First, we give the bilinear
difference equations satisfied by the $\tau$ function
(\ref{dP3tau}). Secondly, we show that eq.(\ref{dP3-2})
follows from the bilinear difference equations.

\subsection{Bilinear Difference Equations}

The bilinear difference equations for the $\tau$ function
(\ref{dP3tau}) are given as
\begin{eqnarray}
&\tau_{N+1}^{\nu}(n)~{}\tau_N^{\nu+1}(n+1)
-q^{-\nu-N}\tau_{N+1}^{\nu}(n+1)~{}\tau_N^{\nu+1}(n)
\nonumber\\
&\quad=-(1-q)q^{n+2N}\tau_{N+1}^{\nu+1}(n)~{}\tau_N^{\nu}(n+1)\ ,
\label{bleq1}\\
&\ \nonumber\\
&\tau_{N+1}^{\nu+1}(n)~{}\tau_N^{\nu}(n+1)-q^{\nu-N+1}
\tau_{N+1}^{\nu+1}(n+1)~{}\tau_N^{\nu}(n)
\nonumber\\
&\quad =(1-q)q^{n+2N}\tau_{N+1}^{\nu}(n)~{}\tau_N^{\nu+1}(n+1)\ ,
\label{bleq2}\\
&\ \nonumber\\
&\tau_{N+1}^{\nu}(n)~{}\tau_N^{\nu+1}(n+3)
-q^{-\nu-N}\tau_{N+1}^{\nu}(n+1)~{}\tau_N^{\nu+1}(n+2)
\nonumber\\
&\quad =-(1-q)q^n\tau_{N+1}^{\nu+1}(n)~{}\tau_N^{\nu}(n+3)\ ,
\label{bleq3}\\
&\ \nonumber\\
&\tau_{N+1}^{\nu+1}(n)~{}\tau_N^{\nu}(n+3)
-q^{\nu-N+1}\tau_{N+1}^{\nu+1}(n+1)~{}\tau_N^{\nu}(n+2)
\nonumber\\
&\quad =(1-q)q^n\tau_{N+1}^{\nu}(n)~{}\tau_N^{\nu+1}(n+3)\ ,
\label{bleq4}\\
&\ \nonumber\\
&\tau_{N+1}^{\nu}(n+2)~{}\tau_N^{\nu+1}(n+1)-q^{2N}(q^{\nu-N}+q^{-\nu+N})
\tau_{N+1}^{\nu}(n+1)~{}\tau_{N}^{\nu+1}(n+2)
\nonumber\\
&+q^{4N}\{ 1+(1-q)^2q^{2n}\} \tau_{N+1}^{\nu}(n)~{}\tau_N^{\nu+1}(n+3)=0\ .
\label{bleq5}
\end{eqnarray}
In the following, we give a detail of the proof that
eqs.(\ref{bleq1})-(\ref{bleq5})
hold for the $\tau$ function (\ref{dP3tau}).\par

\noindent {\bf (1) Difference Formulas}

The Pl\"ucker relations are the quadratic
identities for determinants some of whose columns are shifted.
Bilinear equations are obtained from the relations. Therefore
the first step is to construct several formulas which relate
the ``shifted determinants''.
For usual integrable systems, such as
the KP, Toda lattices and discrete KP, the entries of their
determinant representation of $\tau$ functions
satisfy quite simple linear differential or difference
equations, and hence we can easily construct the differential or
difference formulas for the $\tau$ functions
(see, for example, \ \cite{Sato,Wronski,DKP,RT}).
On the other hand, in this case, the entries of the $\tau$ functions satisfy
linear, but variable-coefficient difference equations as shown
below, which makes the calculations somewhat technical.

We now consider the $\tau$ function (\ref{dP3tau})
provided that the element $J_{\nu}(n)$ satisfies
the contiguity relations (\ref{contig1}) and (\ref{contig2}).
Let us introduce a notation for simplicity,
\begin{equation}
\tau_N^{\nu}(n)=\vert 0_{\nu},1_{\nu},\cdots , N-2_{\nu}, N-1_{\nu}\vert
\ ,\label{F-N1}
\end{equation}
where $k_{\nu}$ stands for the column which begins with $J_{\nu}(n+k)$,
\begin{equation}
k_\nu=\pmatrix{
 J_\nu(n+k)   \cr
 J_\nu(n+k+2) \cr
 J_\nu(n+k+4) \cr
 \vdots}\ .
\label{knu}
\end{equation}
Here the height of the column $k_\nu$ is $N$.
However in the following, we use the same symbol $k_\nu$
for the bigger determinant $\tau_{N+1}^\nu(n)$
because there is no possibility of confusion.
So the size of $k_\nu$ should be read appropriately
case by case.
By adding $(i+1)$-th column
multiplied by $-q^{-\nu}$ to $i$-th column and using eq.(\ref{contig1})
for $i=1\sim (N-1)$, we have
\begin{eqnarray}
\tau_N^{\nu}(n)&=&\{ -(1-q)\}^{N-1}q^{ (2n+N-2)(N-1)\over
2}\nonumber\\
&\times&\left\vert\matrix{
J_{\nu+1}(n) &\cdots &J_{\nu+1}(n+N-2)&J_{\nu}(n+N-1)\cr
q^2J_{\nu+1}(n+2)&\cdots &q^2J_{\nu+1}(n+N)&J_{\nu}(n+N+1)\cr
\vdots     &\cdots &\vdots&\vdots\cr
q^{2N-2}J_{\nu+1}(n+2N-2) &\cdots&q^{2N-2}J_{\nu+1}(n+3N-4)&J_{\nu}
(n+3N-3)\cr}\right\vert
\label{calc1}
\\
&=&\{ -(1-q)\}^{N-1}q^{(2n+3N-2)(N-1)\over
2}\nonumber\\
&\times&\left\vert\matrix{
J_{\nu+1}(n) &\cdots &J_{\nu+1}(n+N-2)&J_{\nu}(n+N-1)\cr
J_{\nu+1}(n+2)&\cdots &J_{\nu+1}(n+N)&q^{-2}J_{\nu}(n+N+1)\cr
\vdots     &\cdots &\vdots&\vdots\cr
J_{\nu+1}(n+2N-2) &\cdots&J_{\nu+1}(n+3N-4)&q^{-2N+2}J_{\nu}
(n+3N-3)\cr}\right\vert
\nonumber\\
&=&\{ -(1-q)\}^{N-1}q^{(2n+3N-2)(N-1)\over 2}\quad\vert 0_{\nu+1},\cdots ,
N-2_{\nu+1}, N-1_{\nu}^\prime\vert\ ,\nonumber
\end{eqnarray}
where $k_{\nu}^\prime$ stands for the column,
\begin{equation}
k_{\nu}^\prime=\pmatrix{ J_{\nu}(n+k)\cr q^{-2}J_{\nu}(n+k+2)\cr
q^{-4}J_{\nu}(n+k+4)\cr \vdots\cr}\ .
\label{knuprime}
\end{equation}
Hence we obtain a difference formula,
\begin{equation}
\vert 0_{\nu},\cdots , N-2_{\nu}, N-1_{\nu-1}^\prime\vert
=\{-(1-q)\}^{-N+1}q^{-{(2n+3N-2)(N-1)\over 2}}\quad\tau_N^{\nu-1}(n).
\label{shift1}
\end{equation}

Let us construct another difference formula.
In eq.(\ref{calc1}), multiplying $N$-th column by $q^{-\nu}$,
adding $(N-1)$-th column multiplied by $-(1-q)q^{n+N-2}$ to $N$-th column
and using eq.(\ref{contig1}),
we have
\begin{eqnarray*}
q^{-\nu}\tau_N^{\nu}(n)&=&\{-(1-q)\}^{N-1}q^{(2n+N-2)(N-1)\over 2}
\\
&\times&
\left\vert\matrix{
J_{\nu+1}(n) &\cdots &J_{\nu+1}(n+N-2)&J_{\nu}(n+N-2)\cr
q^2J_{\nu+1}(n+2)&\cdots &q^2J_{\nu+1}(n+N)&J_{\nu}(n+N)\cr
\vdots     &\cdots &\vdots&\vdots\cr
q^{2N-2}J_{\nu+1}(n+2N-2) &\cdots&q^{2N-2}J_{\nu+1}(n+3N-4)&J_{\nu}
(n+3N-4)\cr}\right\vert
\\
&=&\{ -(1-q)\}^{N-1}q^{(2n+3N-2)(N-1)\over 2}
\\
&\times&
\left\vert\matrix{
J_{\nu+1}(n) &\cdots &J_{\nu+1}(n+N-2)&J_{\nu}(n+N-2)\cr
J_{\nu+1}(n+2)&\cdots &J_{\nu+1}(n+N)&q^{-2}J_{\nu}(n+N)\cr
\vdots     &\cdots &\vdots&\vdots\cr
J_{\nu+1}(n+2N-2) &\cdots&J_{\nu+1}(n+3N-4)&q^{-2N+2}J_{\nu}
(n+3N-4)\cr}\right\vert
\\
&=& \{ -(1-q)\}^{N-1} q^{(2n+3N-2)(N-1)\over 2}
\ \vert 0_{\nu+1}, \cdots ,N-2_{\nu+1}, N-2_{\nu}^\prime\vert .
\end{eqnarray*}
Hence we get a difference formula,
\begin{equation}
\vert 0_{\nu},\cdots ,N-2_{\nu},N-2_{\nu-1}^\prime\vert
=\{-(1-q)\}^{-N+1}q^{-{(2n+3N-2)(N-1)\over 2}-\nu+1}\quad \tau_N^{\nu-1}(n)
\ .\label{shift2}
\end{equation}
Similar calculations give the following formulas by using eq.(\ref{contig2})
instead of (\ref{contig1}),
\begin{equation}
\vert 0_{\nu},\cdots , N-2_{\nu}, N-1_{\nu+1}^\prime\vert
=(1-q)^{-N+1}q^{-{(2n+3N-2)(N-1)\over 2}}\quad\tau_N^{\nu+1}(n)\ ,
\label{shift3}
\end{equation}
\begin{equation}
\vert 0_{\nu},\cdots ,N-2_{\nu},N-2_{\nu+1}^\prime\vert
=(1-q)^{-N+1}q^{-{(2n+3N-2)(N-1)\over 2}+\nu+1}\quad \tau_N^{\nu+1}
(n)\ .\label{shift4}
\end{equation}

So far we have constructed the difference formulas of $\tau$ for
``horizontal shifts''.
Next we consider formulas
for ``vertical shifts''(shifts of rows in eq.(\ref{dP3tau})).
For notational convenience, we take the transpose
of $\tau$,
\begin{eqnarray}
\tau_N^{\nu}(n)&=&\left\vert\matrix{
J_{\nu}(n) &J_{\nu}(n+2)&\cdots &J_{\nu}(n+2N-2)\cr
J_{\nu}(n+1) &J_{\nu}(n+3)&\cdots &J_{\nu}(n+2N-1)\cr
\vdots     &\vdots &\cdots&\vdots\cr
J_{\nu}(n+N-1) &J_{\nu}(n+N+1)&\cdots &J_{\nu}(n+3N-3)\cr}\right\vert
\nonumber\\
&=&\vert \hat0_\nu,\hat2_\nu,\cdots ,
\widehat{2N-4}_\nu,\widehat{2N-2}_\nu\vert\ .
\label{transpose}
\end{eqnarray}
Here $\hat k_\nu$ stands for the column,
\begin{equation}
\hat k_\nu=\pmatrix{
 J_\nu(n+k)   \cr
 J_\nu(n+k+1) \cr
 J_\nu(n+k+2) \cr
 \vdots}\ .
\label{hknu}
\end{equation}
Using the contiguity relations (\ref{contig1}) and (\ref{contig2}),
we can verify
\begin{eqnarray}
&&\hat2_\nu - q^{2\nu}(1+(1-q)^2q^{2n})\ \hat0_\nu \nonumber\\
&&\ =(1-q)q^{n+3\nu+1}\pmatrix{
 \qquad  (1+q^{-2\nu})  J_{\nu+1}(n+1) \hfill \cr
 q^{-1}\{(1+q^{-2\nu+2})J_{\nu+1}(n+2) - (1-q^2)q^{-\nu-1}J_{\nu+1}(n+1)\} \cr
 q^{-2}\{(1+q^{-2\nu+4})J_{\nu+1}(n+3) - (1-q^4)q^{-\nu-1}J_{\nu+1}(n+2)\} \cr
 \vdots}.
\label{contigh}
\end{eqnarray}
In eq.(\ref{transpose}),
we first add $j$-th column multiplied by $-q^{2\nu}(1+(1-q)^2q^{2n+4j-4})$
to $(j+1)$-th column for $j=(N-1)\sim 1$ and use eq.(\ref{contigh}).
Secondly, we multiply $i$-th row by $q^{i-1}/(1+q^{-2\nu+2(i-1)})$ for $i=1\sim
N$.
Finally adding $i$-th row multiplied by $(1-q^{2i})q^{-\nu-1}/(1+q^{-2\nu+2i})$
to $(i+1)$-th row for $i=1\sim (N-1)$, we obtain
\begin{eqnarray}
&{}&|\hat0_\nu^\prime,\hat1_{\nu+1},\hat3_{\nu+1},\cdots,\widehat{2N-3}_{\nu+1}|
\nonumber \\
&=& \{ (1-q)q^{n+3\nu-1}\}^{-N+1}q^{-{N(N-1)\over 2}}\prod_{k=0}^{N-1}
{\displaystyle 1\over\displaystyle 1+q^{-2\nu+2k}}
\quad\tau_N^\nu(n)\ ,
\label{shift5}
\end{eqnarray}
where
\begin{equation}
\hat{k}_{\nu}^\prime=\pmatrix{K_\nu(n+k)_0\cr K_\nu(n+k)_1\cr K_\nu(n+k)_2\cr
\vdots \cr},
\label{hknup}
\end{equation}
and $K_\nu(n)_i$ is defined by the following recursion relation,
\begin{equation}
\left\{
\begin{array}{lll}
K_{\nu}(n)_i&=&{\displaystyle 1\over\displaystyle 1+q^{-2\nu +2i}}
\bigl( q^iJ_{\nu}(n+i)+(1-q^{2i})q^{-\nu-1}K_{\nu}(n)_{i-1}\bigr),
\quad {\rm for}\ i\geq 1,\\
K_{\nu}(n)_0&=&{\displaystyle 1\over\displaystyle 1+q^{-2\nu}}
J_{\nu}(n)\ .
\end{array}
\right.
\end{equation}

Let us derive another difference formula.
We multiply $1$st column of $\tau_N^\nu(n)$ in eq.(\ref{transpose}) by
$-q^{2\nu}(1+(1-q)^2q^{2n})$
and add $2$nd column to $1$st column.
Then we follow the same procedure as above calculation, that is,
add $j$-th column multiplied by $-q^{2\nu}(1+(1-q)^2q^{2n+4j-4})$
to $(j+1)$-th column for $j=(N-1)\sim 2$,
multiply $i$-th row by $q^{i-1}/(1+q^{-2\nu+2(i-1)})$ for $i=1\sim N$,
and add $i$-th row multiplied by $(1-q^{2i})q^{-\nu-1}/(1+q^{-2\nu+2i})$
to $(i+1)$-th row for $i=1\sim (N-1)$.
Then we obtain
\begin{eqnarray}
&{}&|\hat1_{\nu+1},\hat2_\nu^\prime,\hat3_{\nu+1},\cdots,\widehat{2N-3}_{\nu+1}|
\nonumber \\
&=& \{ (1-q)q^{n+3\nu-1}\}^{-N+1}q^{-{N(N-1)\over 2}}\prod_{k=0}^{N-1}
{\displaystyle 1\over\displaystyle 1+q^{-2\nu+2k}} \nonumber\\
&\times& (-q^{2\nu})\{ 1+(1-q)^2q^{2n}\}\quad \tau_N^\nu(n)\ .\label{shift6}
\end{eqnarray}

We have constructed six difference formulas (\ref{shift1}),
(\ref{shift2}), (\ref{shift3}), (\ref{shift4}), (\ref{shift5})
and (\ref{shift6}).
In the following, using these difference formulas for $\tau$ function,
we prove that the Pl\"ucker relations reduce to the bilinear difference
equations
(\ref{bleq1})-(\ref{bleq5}).

\noindent {\bf (2) Bilinear Identities}\par
\noindent {\bf A. Equations (\ref{bleq1}) and (\ref{bleq2})}

First we derive eq.(\ref{bleq1}). Consider the identity of determinant,
\begin{equation}
0=\left\vert\matrix{
 \matrix{0_\nu &1_\nu &\cdots &N-1_\nu} &\vbl{4} &\matrix{\hbox{\O}}
&\vbl{4} &N_\nu &N_{\nu-1}^\prime &\phi_1 \cr
 \multispan{7}\hblfil \cr
 \matrix{\hbox{\O}} &\vbl{4} &\matrix{1_\nu & \cdots & N-1_\nu}
&\vbl{4} &N_\nu &N_{\nu-1}^\prime &\phi_1
}\right\vert\ ,
\end{equation}
where $k_\nu$ and $k_\nu^\prime$ denote columns
defined in eqs.(\ref{knu}) and (\ref{knuprime}),
\O\ means an empty matrix of relevant size and
\begin{equation}
\left.\phi_1 = \pmatrix{ 0\cr \vdots\cr 0\cr 1\cr}\right\}N+1\ .
\label{phi1}
\end{equation}
Applying the Laplace expansion on the right-hand side,
we have
\begin{eqnarray}
0&=&\vert 0_\nu,\cdots , N-1_\nu,N_{\nu-1}^\prime\vert\times
   \vert 1_\nu,\cdots , N-1_\nu,N_\nu,\phi_1\vert\nonumber\\
 &-&\vert 1_\nu,\cdots , N-1_\nu,N_\nu,N_{\nu-1}^\prime \vert\times
   \vert 0_\nu,\cdots , N-1_\nu,\phi_1\vert\nonumber\\
 &-&\vert 0_\nu,\cdots , N-1_\nu,N_\nu\vert\times
   \vert 1_\nu,\cdots , N-1_\nu,N_{\nu-1}^\prime ,\phi_1\vert\ .
\label{Pl1}
\end{eqnarray}
We note that eq.(\ref{Pl1}) is nothing but
one of the Pl\"ucker relations.
This identity is rewritten in terms of $\tau$ by using eqs.
(\ref{F-N1}), (\ref{shift1}) and (\ref{shift2}) as
\begin{eqnarray}
&\ &\tau_{N+1}^{\nu-1}(n)~{}\tau_N^{\nu}(n+1)-q^{-\nu-N+1}
\tau_{N+1}^{\nu-1}(n+1)~{}\tau_N^{\nu}(n)\nonumber\\
&\ &\quad=-(1-q)q^{n+2N}\tau_{N+1}^{\nu}(n)~{}\tau_N^{\nu-1}(n+1)\ ,
\end{eqnarray}
which is essentially the same as eq.(\ref{bleq1}).
Equation (\ref{bleq2}) is derived from the
same identity as eq.(\ref{Pl1}) except that $N_{\nu-1}^\prime$ is
replaced by $N_{\nu+1}^\prime$
and that eqs.(\ref{shift3}) and (\ref{shift4}) are used instead of
eqs.(\ref{shift1}) and (\ref{shift2}).

\noindent {\bf B. Equations (\ref{bleq3}) and (\ref{bleq4})}

We next derive eq.(\ref{bleq3}).
Replacing $\phi_1$ in eq.(\ref{Pl1}) by
\begin{equation}
\phi_2 = \pmatrix{1\cr 0 \cr\vdots\cr 0\cr}\ ,
\end{equation}
we have
\begin{eqnarray}
0&=&\vert 0_\nu,\cdots , N-1_\nu,N_{\nu-1}^\prime\vert\times
   \vert 1_\nu,\cdots , N-1_\nu,N_\nu,\phi_2\vert\nonumber\\
 &-&\vert 1_\nu,\cdots , N-1_\nu,N_\nu,N_{\nu-1}^\prime \vert\times
   \vert 0_\nu,\cdots , N-1_\nu,\phi_2\vert\nonumber\\
 &-&\vert 0_\nu,\cdots , N-1_\nu,N_\nu\vert\times
   \vert 1_\nu,\cdots , N-1_\nu,N_{\nu-1}^\prime ,\phi_2\vert\ ,
\label{id4}
\end{eqnarray}
which can be rewritten in terms of $\tau$ again.
Here we give notes on
the treatise of the determinant including the column $\phi_2$.
As an example, let us consider $\vert 1_\nu,\cdots , N-1_\nu, N_{\nu-1}^\prime
, \phi_2
\vert $. Expanding by $(N+1)$-th column, we have
\begin{eqnarray}
&\ &\vert 1_\nu,\cdots , N-1_\nu, N_{\nu-1}^\prime , \phi_2\vert
\nonumber\\
&=& \left\vert\matrix{
J_{\nu}(n+1) & \cdots & J_{\nu}(n+N-1) & J_{\nu-1}(n+N)&1\cr
J_{\nu}(n+3) & \cdots & J_{\nu}(n+N+1) & q^{-2}J_{\nu-1}(n+N+2)&0\cr
\vdots &\cdots &\vdots &\vdots &\vdots\cr
J_{\nu}(n+2N+1) & \cdots & J_{\nu}(n+3N-1) &q^{-2N}J_{\nu-1}(n+3N)&0\cr}
\right\vert\nonumber\\
&=&(-1)^N\left\vert\matrix{
J_{\nu}(n+3)&\cdots &J_{\nu}(n+N+1)&q^{-2}J_{\nu-1}(n+N+2)\cr
\vdots &\cdots &\vdots &\vdots \cr
J_{\nu}(n+2N+1)&\cdots &J_{\nu}(n+3N-1)&q^{-2N}J_{\nu-1}(n+3N)\cr}
\right\vert\nonumber\\
&=&(-1)^Nq^{-2}\quad\vert 3_\nu,4_\nu,\cdots , N+1_\nu,
N+2_{\nu-1}^\prime\vert\ ,
\end{eqnarray}
which indicates that an additional factor $(-1)^Nq^{-2}$ is needed.
Taking this into account, we obtain
\begin{eqnarray}
&{}&\tau_{N+1}^{\nu-1}(n)~{}\tau_N^{\nu}(n+3)-q^{-\nu-N+1}\tau_{N+1}^{\nu-1}
(n+1)~{}\tau_N^{\nu}(n+2)\nonumber\\
&{}&\quad =-(1-q)q^n\tau_{N+1}^{\nu}(n)~{}\tau_N^{\nu-1}(n+3)\ ,
\end{eqnarray}
which is essentially the same as eq.(\ref{bleq3}).
Similarly, eq.(\ref{bleq4})
is derived from the same identity as eq.(\ref{id4}) except that
the column $N_{\nu-1}^\prime$ is replaced by $N_{\nu+1}^\prime$.\par
\noindent {\bf C. Equation (\ref{bleq5})}

Finally we derive eq.(\ref{bleq5}).
We consider the identity of determinant,
\begin{equation}
0=\left\vert\matrix{
 \hat1_{\nu+1} &\vbl{4} &\matrix{\hat2_\nu^\prime &\hat3_{\nu+1} &\cdots
&\widehat{2N-1}_{\nu+1}}
&\vbl{4} &\hbox{\O}
&\vbl{4} &\widehat{2N+1}_{\nu+1} &\phi_1 \cr
\multispan{8}\hblfil\cr
 \hat1_{\nu+1} &\vbl{4} &\hbox{\O}
&\vbl{4} &\matrix{\hat3_{\nu+1} &\cdots &\widehat{2N-1}_{\nu+1}}
&\vbl{4} &\widehat{2N+1}_{\nu+1} &\phi_1\cr}
\right\vert\ ,\label{id5}
\end{equation}
where $\hat k_\nu$, $\hat k_\nu^\prime$ and $\phi_1$ are defined
in eqs.(\ref{hknu}), (\ref{hknup}) and (\ref{phi1}), respectively.
Applying the Laplace expansion to the right-hand side of
eq.(\ref{id5}), we have
\begin{eqnarray}
0&=&|\hat2_\nu^\prime,\hat3_{\nu+1},\cdots,\widehat{2N-1}_{\nu+1},\widehat{2N+1}_{\nu+1}|
    \times
    |\hat1_{\nu+1},\hat3_{\nu+1},\cdots,\widehat{2N-1}_{\nu+1},\phi_1|
\nonumber\\
&+&|\hat1_{\nu+1},\hat2_\nu^\prime,\hat3_{\nu+1},\cdots,\widehat{2N-1}_{\nu+1}|
    \times
    |\hat3_{\nu+1},\cdots,\widehat{2N-1}_{\nu+1},\widehat{2N+1}_{\nu+1},\phi_1|
\nonumber\\
&-&|\hat1_{\nu+1},\hat3_{\nu+1},\cdots,\widehat{2N-1}_{\nu+1},\widehat{2N+1}_{\nu+1}|
    \times
    |\hat2_\nu^\prime,\hat3_{\nu+1},\cdots,\widehat{2N-1}_{\nu+1},\phi_1|\ .
\end{eqnarray}
Using eqs.(\ref{transpose}), (\ref{shift5}) and (\ref{shift6}), we get
\begin{eqnarray}
&&\tau_{N+1}^{\nu}(n+2)~{}\tau_N^{\nu+1}(n+1)
  -q^{2\nu+2N}\{ 1+(1-q)^2q^{2n}\}\tau_{N+1}^{\nu}(n)~{}\tau_N^{\nu+1}(n+3)
\nonumber\\
&&\ =(1-q)q^{n+\nu+N+1}(q^{2N}+q^{2\nu})
  \tau_{N+1}^{\nu+1}(n+1)~{}\tau_N^{\nu}(n+2)\ .\label{bleq6}
\end{eqnarray}
Combining eqs.(\ref{bleq6}) and (\ref{bleq1}), we obtain eq.(\ref{bleq5}).
Thus we have proved that the $\tau$ function (\ref{dP3tau}) really
satisfies the bilinear difference equations (\ref{bleq1})-(\ref{bleq5}).

\subsection{Derivation of dP$_{\rm\romanno3}$ from Bilinear Difference
Equations}

We now show that eq.(\ref{dP3-2}) actually follows from the
bilinear difference equations (\ref{bleq1})-(\ref{bleq5}).
We first note that eq.(\ref{dP3-2}) is factorized as
\begin{equation}
w_{n+1}w_{n-1}=-\frac{\left( q^{2N}w_n-(1-q)^2q^{2n+\nu+7N}\right)
\left( q^{2N}w_n+(1-q)^2q^{2n-\nu+5N-2}\right)}
{\left(w_n+q^{\nu+N}\right)\left(w_n-q^{-\nu+3N}\right) }\ .
\label{dP3fac}
\end{equation}
Let us rewrite the factors in eq.(\ref{dP3fac}).
Equation (\ref{dP3var}) gives
\begin{equation}
w_n = \frac{ \tau_{N+1}^\nu(n+1)\tau_N^{\nu+1}(n)
-q^{\nu+N}\tau_{N+1}^\nu(n)\tau_N^{\nu+1}(n+1)}
{\tau_{N+1}^\nu(n)\tau_N^{\nu+1}(n+1)}\ ,
\end{equation}
which is rewritten by using eq.(\ref{bleq1}) as
\begin{equation}
w_n = (1-q)q^{n+\nu+3N}\frac{\tau_{N+1}^{\nu+1}(n)\tau_N^\nu(n+1)}
{\tau_{N+1}^\nu(n)\tau_N^{\nu+1}(n+1)}\ .\label{fac1}
\end{equation}
By the definition, it is obvious that
\begin{equation}
w_n+q^{\nu+N}=\frac{\tau_{N+1}^\nu(n+1)\tau_N^{\nu+1}(n)}
{\tau_{N+1}^\nu(n)\tau_N^{\nu+1}(n+1)}\ .\label{fac3}
\end{equation}
Using eq.(\ref{bleq5}), we get
\begin{eqnarray}
w_n-q^{-\nu+3N} &= &
 \frac{\tau_{N+1}^\nu(n+1)\tau_N^{\nu+1}(n)
     - (q^{\nu+N}+q^{-\nu+3N})\tau_{N+1}^\nu(n)\tau_N^{\nu+1}(n+1)}
      {\tau_{N+1}^\nu(n)\tau_N^{\nu+1}(n+1)}
 \nonumber\\
 &= & - q^{4N}\{1+(1-q)^2q^{2(n-1)}\}\frac{\tau_{N+1}^\nu(n-1)
\tau_N^{\nu+1}(n+2)}{\tau_{N+1}^\nu(n)\tau_N^{\nu+1}(n+1)}\ .\label{fac4}
\end{eqnarray}
 From eqs.(\ref{fac1})-(\ref{fac4}), we obtain
\begin{eqnarray}
&{}&w_{n+1}w_{n-1}(w_n+q^{\nu+N})(w_n-q^{-\nu+3N})\nonumber\\
&=& - (1-q)^2q^{2n+2\nu+10N} \{ 1+(1-q)^2q^{2(n-1)}\}
 \nonumber\\
&\times&
\frac{\tau_{N+1}^{\nu+1}(n+1)\tau_N^\nu
(n+2)\tau_{N+1}^{\nu+1}(n-1)\tau_N^\nu(n)}
{\tau_{N+1}^{\nu}(n)^2\tau_N^{\nu+1}(n+1)^2}\ .\label{lhs}
\end{eqnarray}

The factor $q^{2N}w_n-(1-q)^2q^{2n+\nu+7N}$ is rewritten
as follows,
\begin{eqnarray}
&{}&q^{2N}w_n-(1-q)^2q^{2n+\nu+7N}\nonumber\\
&=& (1-q)q^{n+\nu+5N}\frac{\tau_{N+1}^{\nu+1}(n)\tau_N^\nu(n+1)-(1-q)q^{n+2N}
\tau_{N+1}^\nu(n)\tau_N^{\nu+1}(n+1)}
{\tau_{N+1}^\nu(n)\tau_N^{\nu+1}(n+1)}
\nonumber\\
&=& (1-q)q^{n+2\nu+4N+1}\frac{\tau_{N+1}^{\nu+1}(n+1)\tau_N^\nu(n)}
{\tau_{N+1}^\nu(n)\tau_N^{\nu+1}(n+1)}\ ,\label{fac5}
\end{eqnarray}
where we have used eq.(\ref{fac1}) in the first line and
eq.(\ref{bleq2}) in the second line, respectively.
Finally the factor $q^{2N}w_n+(1-q)^2q^{2n-\nu+5N-2}$ is
rewritten as
\begin{eqnarray}
&{}&q^{2N}w_n+(1-q)^2q^{2n-\nu+5N-2}\nonumber\\
&=& (1-q)q^{n+\nu+5N}\frac{\tau_{N+1}^{\nu+1}(n)\tau_N^\nu(n+1)
+(1-q)q^{n-2\nu-2}\tau_{N+1}^\nu(n)\tau_N^{\nu+1}(n+1)}
{\tau_{N+1}^\nu(n)\tau_N^{\nu+1}(n+1)}\nonumber\\
&=& (1-q)q^{n+\nu+5N}\nonumber\\
&{}&\hskip10pt
  \times
   \left\{ \left( q^{-\nu+N-1}\tau_{N+1}^{\nu+1}(n-1)\tau_N^\nu(n+2)
         - (1-q)q^{n-\nu+N-2}\tau_{N+1}^\nu(n-1)\tau_N^{\nu+1}(n+2) \right)
   \right.\nonumber\\
&{}&\hskip20pt
   \left. + (1-q)q^{n-\nu+N-2}
           \left( \tau_{N+1}^\nu(n-1)\tau_N^{\nu+1}(n+2)
            + (1-q)q^{n-1}\tau_{N+1}^{\nu+1}(n-1)\tau_N^\nu(n+2) \right)
   \right\}\nonumber\\
&{}&\hskip20pt
   /\tau_{N+1}^\nu(n)\tau_N^{\nu+1}(n+1)
  \nonumber\\
&=&(1-q)q^{n+6N-1}\{ 1+(1-q)^2q^{2(n-1)}\}
   \frac{\tau_{N+1}^{\nu+1}(n-1)\tau_N^\nu(n+2)}
        {\tau_{N+1}^\nu(n)\tau_N^{\nu+1}(n+1)}
\ ,\label{fac6}
\end{eqnarray}
where we have used eq.(\ref{fac1}) in the first line and
eqs.(\ref{bleq3}) and (\ref{bleq4}) in the second line, respectively.
 From eqs.(\ref{fac5}) and
(\ref{fac6}), we get
\begin{eqnarray}
&{}&-\left( q^{2N}w_n-(1-q)^2q^{2n+\nu+7N}\right)
\left( q^{2N}w_n+(1-q)^2q^{2n-\nu+5N-2}\right)\nonumber\\
&=&-(1-q)^2q^{2n+2\nu+10N} \{ 1+(1-q)^2q^{2(n-1)}\}
\nonumber\\
&\times&
\frac{\tau_{N+1}^{\nu+1}(n+1)\tau_N^\nu(n)
      \tau_{N+1}^{\nu+1}(n-1)\tau_N^\nu (n+2)}
{\tau_{N+1}^{\nu}(n)^2\tau_N^{\nu+1}(n+1)^2}\ ,
\end{eqnarray}
which is the same as eq.(\ref{lhs}).
Thus we have proved that the dP$_{\rm\romanno3}$
eq.(\ref{dP3fac}) follows from the bilinear difference equations
(\ref{bleq1})-(\ref{bleq5}).

\section{$q$-Difference Analogue of P$_{\rm III}$}
We have discussed the solution of dP$_{\rm\romanno3}$,
and shown that it is expressed by the discrete Bessel function.
It should be noted, however, that eq.(\ref{dBessel2})
is essentially the same as
the $q$-Bessel equation,
\begin{equation}
J_{\nu}(q^2x)-(q^{\nu}+q^{-\nu})J_{\nu}(qx)+\{ 1+(1-q)^2x^2\}
J_{\nu}(x)=0\ .\label{q-Bessel}
\end{equation}
Actually, eq.(\ref{dBessel2}) yields eq.(\ref{q-Bessel})
by the replacement $q^n\rightarrow x$.
This fact implies that the dP$_{\rm\romanno3}$ is
a $q$-discrete system rather
than a discrete system in essence. More precisely, let us consider
the following equation instead of the dP$_{\rm\romanno3}$ (\ref{dP3-2}),
\begin{equation}
w(qx)w(q^{-1}x)=\frac{\alpha w(x)^2
+\beta x^2 w(x)+\gamma x^4}
{w(x)^2+\delta w(x)+\alpha}\ .\label{qP3}
\end{equation}
The solutions of eq.(\ref{qP3}), with the parameters being
the same as eq.(\ref{dP3par}), are given by
\begin{equation}
w(x)={\displaystyle \tau_{N+1}^{\nu}(qx)\tau_N^{\nu+1}(x)\over
\displaystyle \tau_{N+1}^{\nu}(x)\tau_N^{\nu+1}(qx)}-q^{\nu+N}\ ,
\label{qP3var}
\end{equation}
and the $\tau$ function $\tau_N^\nu(x)$ is expressed as
\begin{equation}
\tau_N^{\nu}(x)=\left\vert\matrix{
J_{\nu}(x)        &J_{\nu}(qx) &\cdots &J_{\nu}(q^{N-1}x) \cr
J_{\nu}(q^2x)     &J_{\nu}(q^3x)&\cdots &J_{\nu}(q^{N+1}x)\cr
\vdots            &\vdots &\cdots&\vdots\cr
J_{\nu}(q^{2N-2}x)&J_{\nu}(q^{2N-1}x)&\cdots &J_{\nu}(q^{3N-3}x)\cr}\right
\vert\ .\label{qP3tau}
\end{equation}
Here $J_{\nu}(x)$ is the $q$-Bessel function of degree $\nu$. \par
We next mention briefly the Lax pair for eq.(\ref{qP3}). The auxiliary linear
system for dP$_{\rm\romanno3}$ is given by\cite{Iso,dP3}
\begin{eqnarray}
\Phi_n(qh)&=&L_n(h)\Phi_n(h)\ ,\\
\Phi_{n+1}(h)&=&M_n(h)\Phi_n(h)\ ,
\end{eqnarray}
whose compatibility condition is written as
\begin{equation}
M_n(qh)L_n(h)=L_{n+1}(h)M_n(h)\ .\label{dP3Lax}
\end{equation}
If we choose suitable $4\times 4$ matrices as $L$ and $M$\cite{dP3},
eq.(\ref{dP3Lax}) yields dP$_{\rm\romanno3}$ (\ref{dP3-2}).
The linear system for eq.(\ref{qP3}) is written
as $q$-difference system both in $x$ and $h$,
\begin{eqnarray}
\Phi(x;qh)&=&L(x;h)\Phi(x;h)\ ,\\
\Phi(qx;h)&=&M(x;h)\Phi(x;h) ,
\end{eqnarray}
whose compatibility condition yields
\begin{equation}
M(x;qh)L(x;h)=L(qx;h)M(x;h)\label{qLax}\ .
\end{equation}
Let the matrices $L$ and $M$ be
\begin{equation}
L(x;h)=\pmatrix{
\kappa & \frac{\dis c}{\dis w(x)}+\kappa & \frac{\dis c}{\dis w(x)}&0\cr
0 & \frac{\dis q^2\ld}{\dis x^2} & q^2\frac{\dis w(q^{-1}x)+\ld}{\dis x^2}
& q^2\frac{\dis w(q^{-1}x)}{\dis x^2}\cr
\frac{\dis w(x)}{\dis h^4x^2} & 0 & \mu & \frac{\dis w(x)}{\dis x^2}+\mu\cr
\frac{\dis 1}{\dis h^4}\left( \frac{\dis c}{\dis q^4w(q^{-1}x)}
+\frac{\dis \ld}{\dis x^2}\right)&
\frac{\dis c}{\dis h^4q^4 w(q^{-1}x)} & 0 &\frac{\dis \ld}{\dis x^2}\cr}\ ,
\end{equation}
\begin{equation}
M(x;h)=\pmatrix{
\frac{\dis (\kappa x^2-q^2\ld) w(x)}{\dis q^2\ld w(x)+cx^2}&
x^2\frac{\dis \kappa w(x)+c}{\dis q^2\ld w(x)+cx^2}& 0 & 0\cr
0 & 0 & 1& 0\cr
0 & 0& \frac{\dis \mu x^2-\ld}{\dis w(x)+\ld} &
\frac{\dis w(x)+\mu x^2}{\dis w(x)+\ld}\cr
\frac{\dis 1}{\dis h^4}& 0& 0& 0\cr}\ ,
\end{equation}
respectively, where $c$, $\kappa$, $\ld$ and $\mu$ are arbitrary
constants. Then we have from eq.(\ref{qLax}),
\begin{equation}
w(qx)w(q^{-1}x)=\frac{c~(q^2\ld w(x)+c x^2)(w(x)+\mu x^2)}
{q^2(\kappa w(x)+c)(w(x)+\ld)}\ ,
\end{equation}
which is essentially the same as eq.(\ref{qP3}).

Finally, we comment on the limit to P$_{\rm\romanno3}$.
We put $w=(1-q)xu$ and take the limit $q\rightarrow 1$.
Then eq.(\ref{qP3}) with the parameters (\ref{dP3par})
reduces to the P$_{\rm\romanno3}$ eq.(\ref{P3})
with the parameters (\ref{P3par}).
Moreover, the $\tau$ function (\ref{qP3tau})
reduces to that of P$_{\rm\romanno3}$ ({\ref{P3tau})
up to a trivial factor and the dependent variable transformation
(\ref{qP3var}) yields eq.(\ref{P3var}). \par
In summary,
\begin{enumerate}
\item Equation (\ref{qP3}) reduces to the P$_{\rm\romanno3}$ equation in the
limit
$q\rightarrow 1$.
\item Equation (\ref{qP3}) has particular solutions
expressed by the determinant whose entries are given by
the $q$-Bessel functions.
\item The auxiliary linear system and the Lax equation
for eq.(\ref{qP3}) are written as $q$-difference
equations.
\end{enumerate}
These facts imply that eq.(\ref{qP3}) is one candidate for
the $q$-difference analogue of P$_{\rm\romanno3}$.

\section{Concluding Remarks}
In this paper, we have investigated the solutions of special function
type for the dP$_{\rm III}$ equation, and
shown that they are expressed in terms of the Casorati
determinants of the discrete analogue of the Bessel functions.
The solutions are reduced to the Wronskian solutions
of the P$_{\rm III}$ in the continuous limit.
Hence we conclude that the dP$_{\rm III}$ is a good discretization
of the P$_{\rm III}$.

The discrete Bessel functions are transformed to the $q$-Bessel
functions by a simple independent variable transformation.
Based on this observation, we have presented a $q$-difference
analogue of the P$_{\rm III}$, whose particular solutions are
expressed by the Wronski-type determinants of the
$q$-Bessel functions. Moreover,  we have mentioned that
the Lax equation and its auxiliary linear problem
are written as $q$-difference equations.

We expected that the $\tau$ function of the dP$_{\rm III}$ satisfies
the bilinear form of the $q$-cylindrical Toda molecule
equation which we proposed in ref.\cite{qToda}
since that of the P$_{\rm III}$ satisfies the cylindrical Toda molecule
equation\cite{Okamoto}. However, we found it is not the case because of
the asymmetry of structure of the $\tau$ function.
Actually the $\tau$ function of the dP$_{\rm III}$ satisfies
a slightly modified version of the $q$-cylindrical Toda molecule
equation which also reduces to the cylindrical Toda molecule
equation in the limit $q\rightarrow 1$.
Similar asymmetric structure of the $\tau$ function
has also been obtained for the solutions of the dP$_{\rm II}$\cite{dP2}.
We do not know yet whether such asymmetry is essential
or not for the discrete Painlev\'e equations.

It is an interesting problem to study whether
the other discrete Painlev\'e equations
proposed in refs.\ \cite{dPreview1,dPreview2,Iso,Miura,dP4,dP3} admit similar
determinant solutions or not.
The existence of rational solutions together with
their determinant structures is also an interesting subject to
be investigated.

Acknowledgment:
We are grateful to Prof. Grammaticos and Prof. Ramani
for fruitful discussions and useful comments.

\end{document}